\documentclass[aps,prl,reprint]{revtex4-1}
\usepackage{epsfig}
\usepackage{subfigure}
\usepackage{dcolumn}
\usepackage{graphicx,graphics}
\usepackage{amsmath}
\usepackage{amssymb}
\usepackage{bm}
\usepackage{color}
\usepackage[utf8]{inputenc}
\usepackage{pifont}
\usepackage[colorlinks,linkcolor={blue},citecolor={blue},urlcolor={blue}]{hyperref}
\usepackage{mathtools}
\usepackage{booktabs}
\usepackage{float}
\newcommand{\Giret}{ {\rm G}^{{\rm R}}}
\newcommand{\Giadv}{ {\rm G}^{{\rm A}}}

\newcommand{\Giless}{ {\rm G}^{ <}}

\newcommand{\Sigl}{ \Sigma^{ <}}

\newcommand{\bk}{{\bm k}}

\newcommand{\ms}{m^{*}}

\begin{document}

\title{Floquet Control of Indirect Exchange Interaction in Periodically Driven Two-Dimensional Electron Systems}

\author{Mahmoud M. Asmar}
\affiliation{Department of Physics and Astronomy, The University of Alabama, Tuscaloosa, AL 35487, USA }
\author{Wang-Kong Tse }
\affiliation{Department of Physics and Astronomy, The University of Alabama, Tuscaloosa, AL 35487, USA }
\date\today

\begin{abstract}
We present a theory for the Ruderman-Kittel-Kasuya-Yosida (RKKY) interaction mediated by a two-dimensional (2D) electron system subjected to periodic driving. This is demonstrated for a heterostructure consisting of two ferromagnets laterally sandwiching the 2D metallic spacer. Our calculations reveal new non-analytic features in the spin susceptibility. For weak light-matter coupling, the RKKY interaction shows oscillations with a period tunable by the light amplitude and frequency. For stronger light-matter coupling, the interaction becomes non-oscillatory and remains purely ferromagnetic.
\end{abstract}
\maketitle

{\it Introduction}.$-$
Heterostructures of magnetic and non-magnetic materials are a pivotal component for the controlled transfer of information between magnetic layers. This transfer is made possible via the magnetic exchange interaction between the magnetic layers' spins. The effective interaction between the magnetic layers is mediated via the conduction electrons of non-magnetic material, and it is known as the Ruderman-Kittel-Kasuya-Yosida (RKKY) interaction~\cite{RKKY1,RKKY2,RKKY3}. The magnetic exchange coupling in these systems oscillates as a function of the non-magnetic spacer thickness with a period given by its Fermi level~\cite{kittelbook}. The envelop of these oscillations decays as a power law that depends on the dimensionality of the spacer material and the conduction electrons' properties~\cite{Graphenerkky,GrapheneRRKY2,RKKYintgraph,MOS2RKKY3,MOS2RKKY2,RKKYRashbaSem}. Hence, the methods of control of the indirect magnetic exchange have relied on the variation of the metallic layer thickness~\cite{LayersRKKY1,LayersRKKY2,Stiles2005,exp1,exp2,exp3,exp4}, implementation of lower dimensional materials hosting exotic fermions~\cite{Graphenerkky,GrapheneRRKY2,RKKYintgraph,MOS2RKKY3,MOS2RKKY2,RKKYRashbaSem} or excitons \cite{Sham1,Sham2,Melo}, and gate voltage variations~\cite{RKKYvolt1,RKKYvolt2,RKKYvolt3,RKKYvolt4}.

An effective tool for the control of quantum systems is time-periodic driving. Periodic irradiation of a solid leads to photo-induced renormalization of the Bloch bands and results in the formation of Floquet-Bloch states. These states can be tuned by the intensity and frequency of light and manifest properties not present in their parent equilibrium system. Periodically irradiated systems have been used in the control of tunneling currents~\cite{FloqTunn,FloqTunn1,FloqTunn2,FloqTunn3,FloqTunn4}, transport properties~\cite{Floqtrans,Floqtrans1,Floqtrans2,Floqtrans3}, bound states~\cite{Floqboud,Floqboud1}, and topological phase transitions~\cite{Floqtop,Floqtop1,Floqtop2,Floqtop3}.

The irradiation of heterostructures of magnetic and non-magnetic materials can provide a heretofore unexplored degree of control for the magnetic exchange coupling in these systems. The key idea for this control is the modification of the carriers mediating the RKKY interaction between the magnets in these heterostructures, which will provide additional control knobs of the exchange interaction. In this work, we theoretically investigate the magnetic exchange in a monochromatically irradiated  magnetic lateral heterostructure (MLH) composed of a left-side ferromagnet ($F_{{\rm L}}$), a two-dimensional electron gas (2DEG), and a right-side ferromagnet ($F_{{\rm R}}$). Using the Keldysh-Floquet formalism we formulate a theory for the non-equilibrium magnetic exchange interaction in MLHs. Our theory predicts that the strong irradiation of the 2DEG leads to a considerable  modification of its Fermi surface properties and consequently the RKKY oscillations in the system. Two main regimes summarize our findings: In the first regime, the system displays RKKY oscillation between ferromagnetic (FM) and anti-ferromagnetic (AFM) coupling with a period that is fully controllable by the frequency and amplitude of the light. In the second regime, we find that the exchange interaction becomes purely ferromagnetic, resembling the exchange interaction in an insulator at equilibrium.

\begin{figure}
  \centering
  \includegraphics[width=\columnwidth]{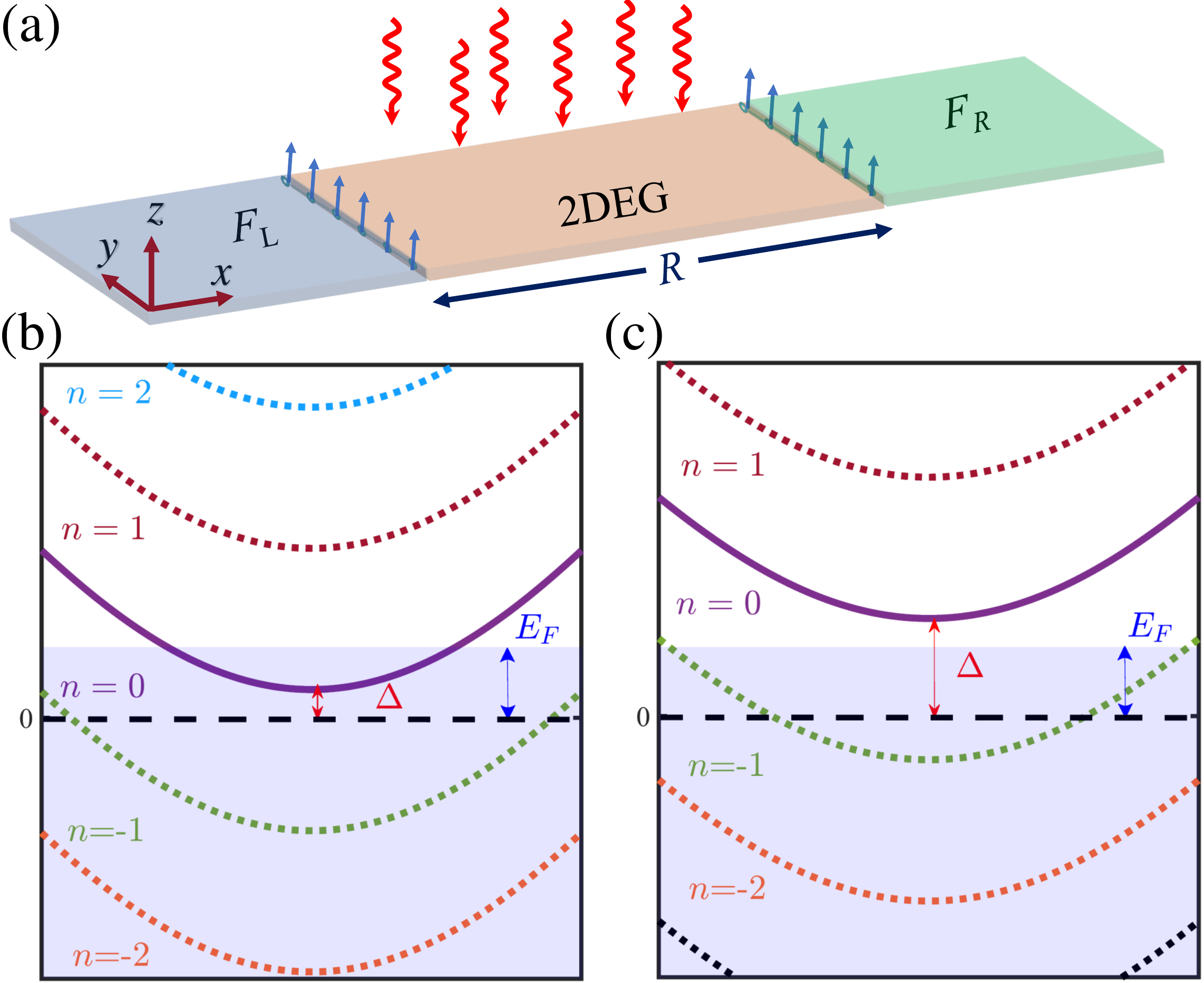}
  \caption{(a) Schematic representation of a MLH composed of left and right ferromagnets, $F_{{\rm L}}$ and $F_{{\rm R}}$, separated by a distance $R$ and laterally sandwiching an irradiated 2DEG. The length of the MLH in the y-direction is assumed  to be much larger than $R$ in the x-direction, and the figure here displays a section of the MLH. (b) and (c) Floquet bands of the irradiated 2DEG. In both (b) and (c) $E_{F}$ is fixed by coupling to a fermion bath, with $E_{F}>\Delta$ in (b) and $E_{F}\le\Delta$ in (c). }\label{fig1}
\end{figure}

{\it Model and Theory}. Our system consists of two metallic ferromagnets ($F_{\rm L}$ and $F_{\rm R}$) adjacent to a monochromatically irradiated 2DEG, as shown in Fig.~\ref{fig1}(a). The ferromagnets are  not irradiated.
At each ferromagnet-2DEG interface, the spins ${\bm S}_{i}$ are located at ${\bm R}_{i}$ and couple
to the 2DEG via the contact potential $\mathcal{V}_{i}({\bm r})=J_{0}\mathcal{S}\cdot{\bm S}_{i}\delta({\bm r}-{\bm R}_{i})$, where $J_{0}$ is the coupling amplitude, $\mathcal{S}=\mu_{{\rm B}}{\bm \sigma}$ is the spin magnetic moment operator of the 2DEG with ${\bm \sigma}$ the Pauli matrices and $\mu_{\rm B}$ the Bohr magneton.
The interaction between a spin of $F_{{\rm L}}$ at ${\bm R}_{i}$ and a spin of $F_{{\rm R}}$ at ${\bm R}_{j}$ is described by the time-dependent exchange  interaction Hamiltonian, $[H_{{\rm int}}(\tau)]_{\mathcal{O},j}=J_{0}\int d {\bm r}\delta({\bm r}-{\bm R}_{j}){\bm S}_{j}\cdot\langle\mathcal{S}({\bm r},\tau)\rangle$, where ${\bm R}_{i}$ can be taken as the origin $\mathcal{O}$.
$\langle\mathcal{S}({\bm r},\tau)\rangle$ is the induced spin magnetic moment in response to ${\bm S}_{i}$ and can be expressed in terms of the non-interacting Green's functions of the irradiated 2DEG.
Summing over $i$ and $j$, we obtain the time-averaged exchange interaction between the two ferromagnets with spin orientations ${\bm S}^{{\rm R}}$ and ${\bm S}^{{\rm L}}$ \cite{supp}
\begin{equation}\label{interexch}
  I(x)=\sum_{\mu=x,y,z}\mathcal{I}_{\mu\mu}\int_{-\infty}^{\infty}dq_{x}e^{iq_{x}x} \chi(q_{x},0)\;,
\end{equation}
where $\mathcal{I}_{\mu\mu}=-(J_{0}\mu_{\rm B})^2S^{{\rm L}}_{\mu}S^{{\rm R}}_{\mu}/(2\pi A_{2{ \rm D}} )$, $A_{2{\rm D}}$ is the 2D unit cell area, and $S^{{\rm L}}_{\mu}$ ($S^{{\rm R}}_{\mu}$) is the $\mu$-projection of the spins in the left (right) ferromagnet. $\chi(q_{x},0)$ is the time-averaged spin susceptibility in the Keldysh-Floquet representation,
\begin{eqnarray}\label{suc}
  &&\chi(q_{x},0)= \frac{2i}{(2\pi)^3}\int_{-\hbar\Omega/2}^{\hbar\Omega/2}d\omega\int d\bk  \\ &&\times\small\sum_{m,n}{\left[\Giret_{nm,\bk+\bm{q}}(\omega)\Giless_{mn,\bk}(\omega)+\Giless_{nm,\bk+\bm{q}}(\omega)\Giadv_{mn,\bk}(\omega)\right]},\nonumber
\end{eqnarray}
where ${\rm G}^{X}_{nm}$ is the matrix component of the non-equilibrium Green's function in the Floquet representation~\cite{Folqthe}, and $X={\rm R, A}$, $<$ denote the retarded, advanced and lesser components, respectively. Note that the right-hand side only depends on $q_x$ through $\bm{q}= (q_x,0)$.

After the continuous light field is turned on and transient dynamics has subsided, the system enters into a
non-equilibrium steady state (NESS) due to energy relaxation. Dissipation is included in our theory through electron tunneling between the 2DEG and a fermion bath \cite{Magnet_Bath} that remains in thermal equilibrium.
When the system has fully relaxed into NESS, time periodicity is restored and Floquet theory \cite{Shirley} applies.
This allows one to determine \cite{supp} the full Green's functions $\rm G$ in terms of the 2DEG's non-interacting Green's functions $\mathcal{G}$ and the self-energies $\Sigma$ using the Keldysh-Floquet  formalism~\cite{Folqthe,Oka_RMP},
\begin{eqnarray}\label{glessint}
 \Giless_{nm,\bk}(\omega)=\sum_{s,l}\Giret_{ns,\bk}(\omega)\Sigl_{sl,\bk}(\omega)\Giadv_{lm,\bk}(\omega)\;.
\end{eqnarray}
$\Sigl_{\bk}(\omega)$ is obtained from integrating over the fermionic bath degrees of freedom. Taking the wide-band approximation for the bath ~\cite{WB}, we get $\Sigl_{nm}(\omega)=2i\Gamma f(\omega+n\hbar\Omega)\delta_{n,m}$, where $ f(\omega)$ is the Fermi distribution function, $\Gamma=\pi |V|^2 \rho$ with $V$ being the 2DEG-bath coupling strength and $\rho$ the fermion bath's density of states. The retarded and advanced Green's functions are given by the Floquet Hamiltonian $H_{F}$,
\begin{equation}\label{Gra}
  {\rm G}_{\bk,mn}^{{\rm R /A}}(\omega)= \left[\left(\omega\pm i\Gamma\right)\delta_{m,n} -H_{F,mn}(\bk)\right]^{-1}\;.
\end{equation}

The 2DEG is taken to have a single parabolic band described by the equilibrium Hamiltonian $H_{0}=\hbar^2k^2/(2\ms)$, where $m^*$ is the effective mass.
In the presence of a circularly polarized (CP) light normally incident onto the 2DEG, the irradiated 2DEG Hamiltonian follows from the minimal substitution $\bm{k} \rightarrow \bm{k}+(e/\hbar)\bm{A}(\tau)$ in $H_{0}$, where $e>0$ is the electron charge, ${\bm A}(\tau)=(E_{0}/\Omega)[\sin(\Omega \tau),\eta \cos(\Omega\tau)]$  is the vector potential of the incident laser, with an electric field amplitude $E_{0}$, frequency $\Omega$ and CP light helicity $\eta=\pm$. Hence, the Schr\"odinger equation of the light-driven system is
\begin{equation}\label{tdepH}
\frac{\hbar^{2}}{2\ms}|\bk+(e/\hbar){\bm A}(\tau)|^2|\psi(\tau)\rangle=i\hbar\partial_{\tau}|\psi(\tau)\rangle\;.
\end{equation}
Due to time periodicity the Floquet-Bloch theorem grants Eq.~(\ref{tdepH}) with solutions of the form $|\psi_{l,\bm{k}}(\tau)\rangle=e^{-i\epsilon_{l, \bm{k}}\tau/\hbar}|\Phi_{l, \bm{k}}(\tau)\rangle$, where, $l \in \mathbb{Z}$ labels the Floquet modes and $\epsilon_{l, \bm{k}}$ is the quasienergy defined modulo $\hbar\Omega$.
The time periodicity of the Floquet state, $|\Phi_{l, \bm{k}}(\tau)\rangle = |\Phi_{l, \bm{k}}(\tau+2\pi/\Omega)\rangle$, enables a Fourier series representation  $|\Phi_{l, \bm{k}}(\tau)\rangle=\sum_{n=-\infty}^{\infty}e^{-in\Omega \tau}|\phi^{n}_{l, \bm{k}}\rangle$. Substituting $|\psi_{l, \bm{k}}(\tau)\rangle$ thus obtained into Eq.~(\ref{tdepH}) casts the original Schr\"odinger equation into an eigenvalue problem of the Floquet Hamiltonian $H_F =H(\tau)-i\hbar\partial_{\tau}$ in the Fourier domain.
For a CP-irradiated 2DEG, $H_{F}$ takes a tridiagonal form
\begin{eqnarray}\label{HFCP}
  H_{F,mn}&&=\left(\frac{\hbar^{2}k^{2}}{2\ms}+\mathcal{V}_{2}-m\hbar\Omega\right)\delta_{m,n}-\\
          &&i\mathcal{V}_{1}\left[\delta_{m+1,n}(k_{x}+i\eta k_{y})-\delta_{m-1,n}(k_{x}-i\eta k_{y})\right], \nonumber
\end{eqnarray}
where  $\mathcal{V}_{1}$ and $\mathcal{V}_{2}$ are defined via $\mathcal{V}_{1}\ms/\hbar^2$$=$$[eE_{0}/(\hbar\Omega)]$, and $2\mathcal{V}_{2}\ms/\hbar^2$$=$$[eE_{0}/(\hbar \Omega)]^2$. The quasienergy eigenvalues of the Floquet Hamiltonian in this case are exactly solvable and given by $\epsilon_{l,k}=\hbar^{2}k^{2}/{(2\ms)}+\mathcal{V}_{2}-l\hbar\Omega$~\cite{supp}.

From the Floquet Hamiltonian Eq.~\eqref{HFCP}, we notice that the light-matter coupling term $\sim \mathcal{V}_{1}k_{x,y}$ is momentum-dependent.  Eq.~\eqref{HFCP} thus implies an undesirable feature that the problem will become dependent on the energy cutoff implicit in the continuum model description of the 2DEG. This is clearly unphysical, and requires a remedy. To obtain an exchange coupling that is independent of the theory's cutoff, we regularize the continuum model by projecting it onto a tight-binding (TB) Hamiltonian that captures the Floquet dynamics of the irradiated low-energy Hamiltonian. To determine the regularized Floquet Hamiltonian, we first construct the general Floquet-TB Hamiltonian for an irradiated square lattice with a lattice constant $a$. Imposing the recovery of the Floquet Hamiltonian in Eq.~\eqref{HFCP} for small momenta, we find a constraint on the coupling strength $\mathcal{A}a\ll 2\sqrt{2}$~\cite{supp} with $\mathcal{A}=eE_{0}/(\hbar \Omega)$
leading to the regularized Hamiltonian
 \begin{eqnarray}\label{tridtb}
 H^{\rm {TB}}_{F,mn}&&=\left(\bar{\epsilon}_{\bk} -m\hbar\Omega\right)\delta_{m,n}\\
           &&-2i{t}J_{1}(a\mathcal{A})\left[\sin(k_{x}a)+i\eta\sin(k_{y}a)\right]\delta_{m,n-1}\nonumber\\
           &&+2i{t}J_{1}(a\mathcal{A})\left[\sin(k_{x}a)-i\eta\sin(k_{y}a)\right]\delta_{m,n+1}\nonumber\;,
\end{eqnarray}
where, $J_{n}$ is the Bessel function of order $n$, $ta^{2}=\hbar^{2}/(2\ms)$, $\pi/a=k_{c}$ determines the bandwidth in the continuum approximation, and quasienergy of the $n=0$ Floquet mode is
\begin{equation}\label{n0eng}
\bar{\epsilon}_{\bk}=2{t}\left\{2-J_{0}(a\mathcal{A})\left[\cos(k_{x}a)+\cos(k_{y}a)\right]\right\}\;.
\end{equation}
\begin{figure}
  \centering
  \includegraphics[width=\columnwidth]{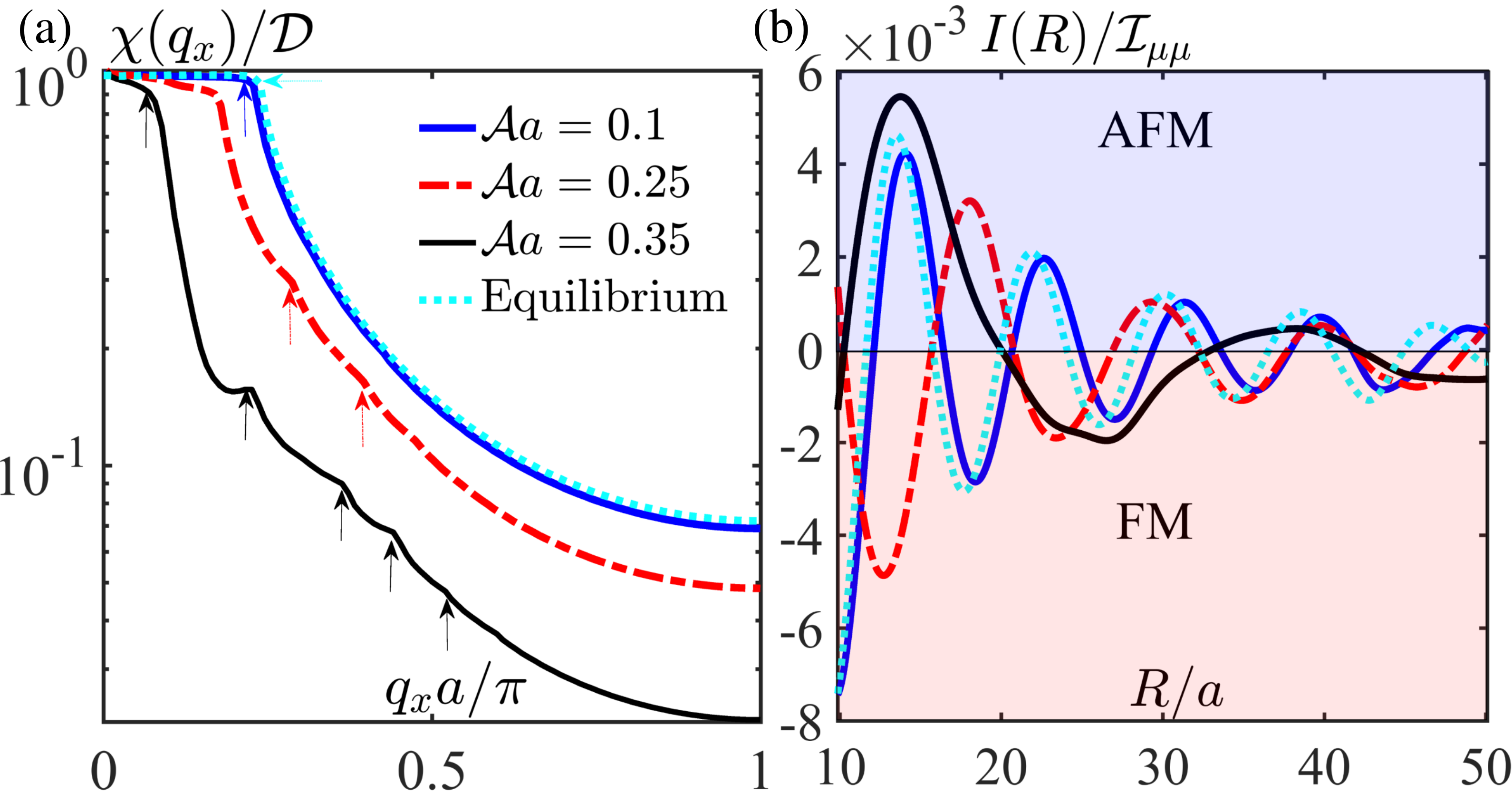}
  \caption{Spin susceptibility and magnetic exchange interaction for the MLH in Fig.~\ref{fig1}(a). The 2DEG is irradiated with a CP light characterized by an energy $\hbar\Omega=0.3$ eV for values of the electric field amplitude in the range $E_{0}=84-300$ MV/m leading to a coupling parameter $\mathcal{A}a=0.1-0.35$. In this regime the coupling of the 2DEG to the fermion bath sets the Fermi level at $E_{F}=140$ meV.
  (a) Spin susceptibility in units of $\mathcal{D}=m/(\pi\hbar^2)$ for the light-matter coupling values $\mathcal{A}a=0.1,0.25,0.35$, in addition to the equilibrium case. The locations of the various Kohn anomalies are indicated by the arrows. (b) Magnetic exchange interaction of the irradiated MLH, $I(R)/\mathcal{I}_{\mu\mu}$ ($\mu=x,y$ or $z$) where $\mathcal{I}_{\mu\mu}$ is given in Eq.~\eqref{interexch} and magnetization directions are assumed to be parallel, for the same values of $\mathcal{A}a$ as in (a). In all our calculations we set $t=1$ eV and $\Gamma=3$ meV.}\label{fig2}
\end{figure}
\begin{figure}
  \centering
  \includegraphics[width=\columnwidth]{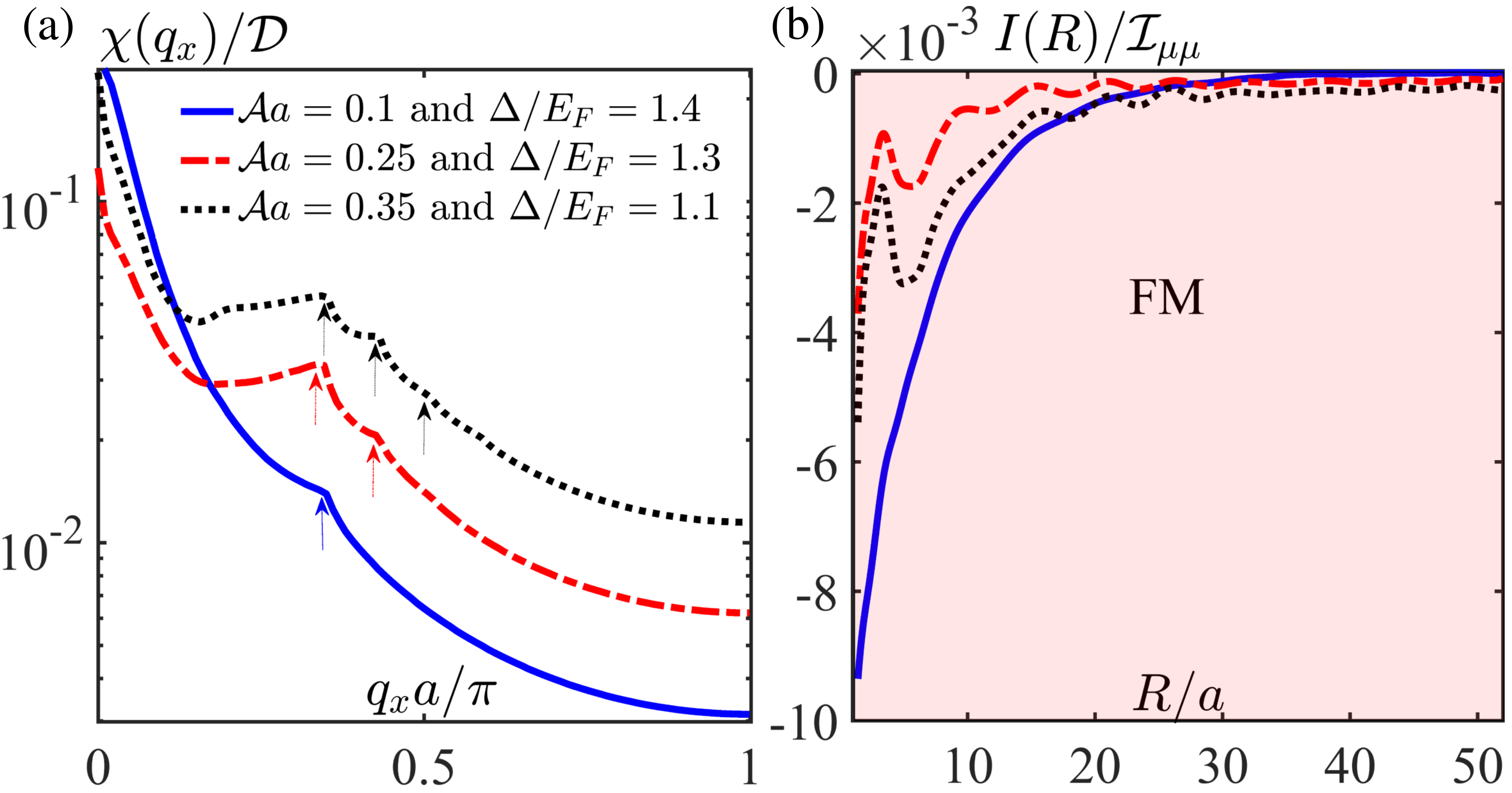}
  \caption{Spin-susceptibility and magnetic exchange interaction of an irradiated MLH. The system is illuminated with $\hbar\Omega=0.3$ eV, and three distinct light-matter coupling amplitudes. (a) Spin-susceptibility for $\mathcal{A}a=0.1$ with $E_{F}=0.72\Delta$, $\mathcal{A}a=0.25$ with $E_{F}=0.77\Delta$, and $\mathcal{A}a=0.35$ with $E_{F}=0.91\Delta$. (b) Corresponding exchange interaction  $I(R)/\mathcal{I}_{\mu\mu}$ ($\mu=x, y$ or $z$) between the ferromagnets with parallel magnetization directions. Other parameters are the same as in Fig.~\ref{fig2}.}\label{fig3}
\end{figure}

The irradiation of the 2DEG in MLH in Fig.~\ref{fig1}(a) leads to a renormalization of its equilibrium band into Eq.~\eqref{n0eng} and the appearance of the photon-dressed Floquet sidebands that are separated by $\hbar\Omega$ [Fig.~\ref{fig1}(b) and (c)]. The  renormalization of the equilibrium dispersion leads to an energy shift of this band, given by $\Delta=\bar{\epsilon}_{\bk=0}=4t[1-J_{0}(a\mathcal{A})]\approx \mathcal{V}_{2}$, where $\mathcal{V}_{2}$ is given by Eq.~\eqref{HFCP} and increases with light-matter coupling. Since the
Fermi level of the 2DEG,
$E_{F}$, is assumed fixed by coupling to the fermion bath we expect two distinct regimes of the magnetic exchange with increasing light-matter coupling: $\Delta<E_{F}$ [Fig.~\ref{fig1}(b)] and $\Delta\ge E_{F}$ [Figs.~\ref{fig1}(c)].

\textit{Oscillatory Magnetic Exchange in MLHs: $\Delta<E_{F}$}.
The RKKY interaction is mediated by
the spin density oscillations induced
by the localized moments of the ferromagnets, which in equilibrium is associated with the non-analytic behavior of the spin susceptibility at $2k_F$, known as Kohn anomaly ~\cite{kohnanomaly}.
Under irradiation, we find that the non-equilibrium spin susceptibility of 2DEG exhibits additional Kohn anomalies, consistent with a recent finding \cite{ono}.
In the following we focus on the effects of irradiation on these Kohn anomalies, the spin susceptibility and the RKKY interaction.

When the light-matter coupling is small  $\mathcal{A}a \ll 1$, $\chi(q_{x})$ displays a single Kohn anomaly at $q_{x}=2k_{F,0}$ resulting from the $n=0$ Floquet band, where $k_{F,n}$ denotes the Fermi wave vector Eq.~\eqref{n0eng} of the $n$th Floquet sideband. Since the light-induced shift $\Delta\ll E_{F}$, the effect of light remains weak and the spin susceptibility $\chi(q_{x})$ resembles that in equilibrium, as seen in Fig.~\ref{fig2}(a) for $\mathcal{A}a=0.1$.
Increasing the light-matter coupling increases the spectral weight of the Floquet sidebands, and this causes the Kohn anomalies associated with the intersection of the Fermi surface with these sidebands to become more prominent in $\chi(q_{x})$.
Hence, for $\mathcal{A}a=0.25$, due to an increase in $\Delta$ the position of the Kohn anomaly at $q_{x}=2k_{F,0}$ is shifted and two additional cusps appear.
The first one, located at $q_{x}=k_{F,0}+k_{F,-1}$, is the nesting vector connecting the adjacent, $n=0$ and $n=-1$, Floquet sidebands (``interband Kohn anomaly'').
The second originates solely from the $n=-1$ Floquet sideband and appears at $q_{x}=2k_{F,-1}$~\cite{supp} (``intraband Kohn anomaly'').
With further increase of light-matter coupling, the $\mathcal{A}a=0.35$ case shows
five clearly visible Kohn anomalies, with the intraband Kohn anomalies occurring at $q_{x}=2k_{F,0}, 2k_{F,-1}, 2k_{F,-2}$ and the interband ones at $q_{x}=k_{F,n}+k_{F,n-1}$ with $n=-1$ and $n=0$.

The RKKY coupling under irradiation in Fig.~\ref{fig2}(b) shows an oscillation period that is increasing with the light-matter coupling $\mathcal{A}a$. The dominant component of the RKKY oscillation period is given by the $n = 0$ Kohn anomaly $q_{x}=2k_{F,0}$. The increase in period with the coupling $\mathcal{A}a$ can be understood from the decreased Fermi energy measured from the band edge as a result of increased $\Delta$ [Fig.~\ref{fig1}(a)]. When light-matter coupling is strong enough and the chemical potential is held constant by coupling to leads, irradiation serves essentially as ``dynamic gating'' by tuning the energy shift $\Delta$  of the band.
The period of the exchange interaction in the irradiated system, $\Lambda(\Omega, E_{0})$, is determined by
\begin{equation}\label{modperiod}
  \Lambda(\Omega, E_{0})=\frac{\pi a}{\sqrt{2}\cos^{-1}\left\{[4t-E_{F}]/[4tJ_{0}(\mathcal{A}a)]\right\}}\;.
\end{equation}
The behaviour described by Eq.~\eqref{modperiod} is valid as long as $\Delta<E_{F}$. On top of the main signature on the oscillation period due to the $n=0$ Floquet band, for large values of the coupling ($\mathcal{A}a=0.25,0.35$) the period of oscillations is also weakly modulated by secondary periods resulting from other intraband and interband Kohn anomalies.

{\it Non-Oscillatory Magnetic Exchange in MLHs: $\Delta\ge E_{F}$}. When the light-matter couping is strong enough such that the energy shift $\Delta$ exceeds the Fermi level set by the fermion bath,
the RKKY interaction undergoes a striking qualitative change. As seen in the quasienergy picture [Fig.~\ref{fig1}(b)], the $n = 0$ Floquet band is now elevated above the Fermi level and as a result its $n = 0$ Kohn anomaly is absent in the susceptibility [Fig.~\ref{fig3}(a)]. However, the Fermi level still intersects all the $n \le -1$ sidebands and one expects their corresponding Kohn anomalies to appear.
For weak coupling $\mathcal{A}a \ll 1$, a single intraband Kohn anomaly is seen due to the $n=-1$ Floquet sideband, as demonstrated in Fig.~\ref{fig3}(a) with $\mathcal{A}a=0.1$. As the coupling strength is increased ($\mathcal{A}a=0.25$), an additional interband Kohn anomaly connecting the $n=-1$ and $n=-2$ sidebands becomes apparent. On further increase of the coupling ($\mathcal{A}a=0.35$), the intraband Kohn anomaly due to the $n=-2$ sideband appears.

In contrast to the previous case $\Delta < E_{F}$, the exchange coupling in the present regime does not oscillate between FM and AFM behaviors as a function of $R$. Instead, it displays a decaying profile with $R$ modulated by weak oscillations and remains in the FM regime for all $R$ values.
This behaviour is a unique manifestation of the strong irradiation effects and is not present in the parent equilibrium system. With the Fermi level below the main $n = 0$ Floquet band, the RKKY coupling  resembles the behavior of
insulators~\cite{ins,ins1,ins2,ins3}, intrinsic semiconductors (\textit{e.g.}, undoped MoS$_{2}$~\cite{MOS2RKKY}) and semimetals (\textit{e.g.}, intrinsic graphene~\cite{RKKYintgraph}), due to coupling through evanescent modes.

\begin{figure}
  \includegraphics[width=\columnwidth]{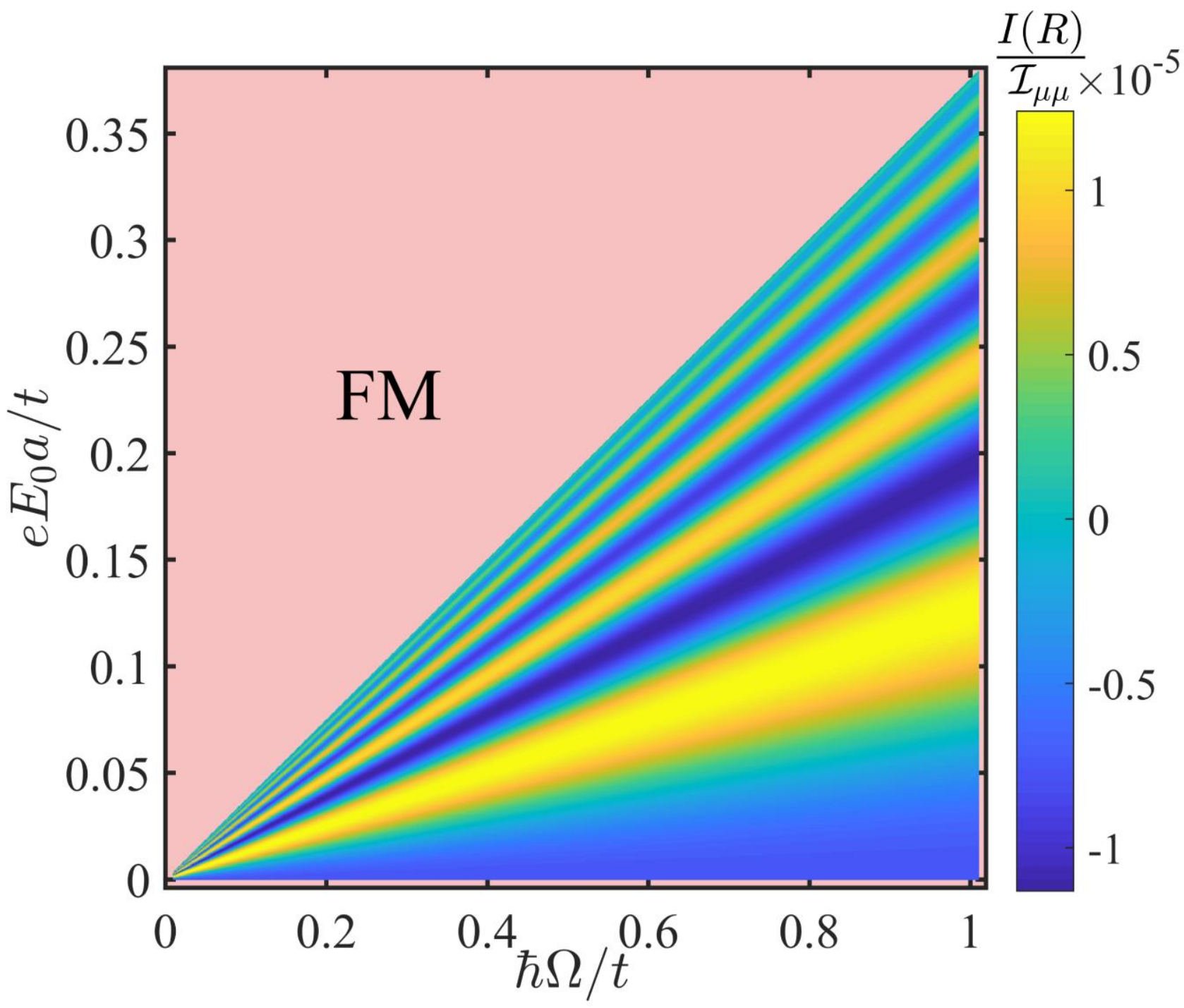}
  \caption{Exchange interaction of the irradiated MLH with a constant spacer width $R=18$ nm and Fermi energy $E_{F}=140$ meV, as a function of light frequency $\Omega$ and field strength $E_{0}$ scaled in their appropriate units. The boundary line between the two regimes is given by $eE_{0}a=\hbar\Omega\sqrt{E_{F}/t}$~\cite{supp}.}\label{fig4}
\end{figure}

For the first regime $\Delta/E_{F}<1$, we can approximately capture the principal asymptotic $k_{F,0}R\gg 1$ behavior of the exchange interaction by considering the dominant $n = 0$ Floquet mode with a shifted Fermi energy $E_{F}-\Delta$, yielding $I(R)\approx-\mathcal{I}_{\mu\mu}$$\mathcal{C}\sin(2\pi R/\Lambda+$$\pi/4)/R^{3/2}$~\cite{supp}, where $\Lambda$ is in Eq.~\eqref{modperiod} and $2\mathcal{C}^{-1}=\sqrt{ta^2\Lambda E_{F}/2}$. Fig.~\ref{fig4} shows this approximate exchange coupling as a function of the laser field amplitude $E_0$ and frequency $\Omega$, showing a wide range of tunability between FM and AFM behaviors at a fixed $R$. When the field amplitude becomes large or frequency becomes small such that the coupling $\mathcal{A}a \ge \sqrt{E_F/t}$, the system enters into the second regime $\Delta/E_{F}>1$ and behaves ferromagnetically. The dominant $n = 0$ Floquet mode follows approximately an exponential decay $I(R)\sim -\mathcal{I}_{\mu\mu}\mathcal{C}'\exp{(-2\kappa_{F,0}R)}/R^{3/2}$ with $\kappa_{F,0}=\pi/|\Lambda|$~\cite{supp}, and $\mathcal{C}'=\sqrt{ta^2\pi E_{F}/(2\kappa_{F,0})}$.

{\it Experimental Realization}. One proposal to realize MLHs is to sandwich a conventional high-mobility GaAs-Al$_x$Ga$_{1-x}$As heterostructure ~\cite{Aoki_exp1,Aoki_exp2,system,system1,system2,system3} laterally between two ferromagnets. A laser spot can be illuminated onto the 2DEG layer without illuminating the ferromagnets. The driving frequency should satisfy $\Gamma \ll \hbar\Omega \ll \Delta_{{\rm sb}}-E_F$ to ensure that quasienergy bands are well-resolved and originate from only the 2DEG's lowest subband, which is separated from the next higher subband by $\Delta_{{\rm sb}}$. We can estimate $\Delta_{{\rm sb}}$ using a square well of width $L_{z}$ to obtain $\Delta_{{\rm sb}}\approx 1.8\times 10^{3}[{\rm eV \AA^{2}}]/L^{2}_{z}$. For typical values of Fermi energy $E_{F}=140$ meV and $L_{z}=60$\AA, $\hbar\Omega \lesssim 360$ meV within the infrared spectrum. We also propose atomically thin 2D systems as suitable platforms to investigate Floquet-driven RKKY interaction.
Recently discovered atomically thin 2D ferromagnets  (\textit{e.g.}, Fe$_{3}$GaTe$_{2}$~\cite{2dmag}) could provide a realization  of MLHs in the atomic scale by laterally  depositing two such ferromagnetic layers on another atomically thin material such as doped graphene. In these MLH setups, the measurement of the exchange coupling can be performed via magneto-resistance oscillations experiments~\cite{magres} or via spin-polarized scanning tunneling microscopy experiments~\cite{chains3}.

\acknowledgments {We thank Woo-Ram Lee and Jalen Cates for useful discussions. This work was supported by the U.S. Department of Energy, Office of Science, Basic Energy Sciences under Early Career Award $\#$DE-SC0019326.}

\bibliographystyle{apsrev4-2}
\bibliography{Refs}

\end{document}